\documentclass{nature}
\usepackage[english]{babel}
\usepackage[T1]{fontenc}
\usepackage{csquotes}
\usepackage[fleqn]{amsmath}
\usepackage{amsmath}
\usepackage{pdfpages}
\usepackage{wasysym}
\usepackage{amsmath}
\usepackage{pdfpages}
\usepackage{multirow}
\usepackage{array}

\graphicspath{ {Figures/} }
\title{Vulnerabilities of Electric Vehicle Battery Packs to Cyberattacks}

\author{Shashank Sripad$^1$, Sekar Kulandaivel$^2$,  Vikram Pande$^1$, Vyas Sekar$^2$ \& Venkatasubramanian Viswanathan$^1$}

\usepackage{graphicx}
\makeatletter
\let\saved@includegraphics\includegraphics
\AtBeginDocument{\let\includegraphics\saved@includegraphics}
\makeatother

\begin{document}

\maketitle

\begin{affiliations}
 \item Department of Mechanical Engineering, Carnegie Mellon University, Pittsburgh, Pennsylvania 15213.
 \item Department of Electrical and Computer Engineering, Carnegie Mellon University, Pittsburgh, Pennsylvania 15213.
\end{affiliations}

\maketitle

\begin{abstract}
Electric Vehicles (EVs), like all modern vehicles, are entirely controlled by electronic devices embedded within networks that are exposed to the threat of cyberattacks. Cyber vulnerabilities are magnified with EVs due to unique risks associated with EV battery packs. Current batteries have well-known issues with specific energy, cost and fire-related safety risks. In this study, we develop a systematic framework to assess the impact of cyberattacks on EVs. While the current focus of automotive cyberattacks is on short-term physical safety, it is crucial to consider long-term cyberattacks that aim to cause financial losses through accrued impact, especially in the context of EVs. Faulty components of battery management systems such as a compromised voltage regulator could lead to cyberattacks that can overdischarge or overcharge the battery. Overdischarge could lead to failures such as internal shorts in the timescale of minutes through cyberattacks that compromise energy-intensive EV subsystems like auxiliary components.  Attacks that overcharge the pack could shorten the lifetime of a new battery pack to less than a year.  Further, such attacks also pose physical safety risks via the triggering of thermal (fire) events.  Attacks on auxiliary components lead to battery drain, which could be up to 20\% of the state-of-charge per hour.  Lastly, we develop a heuristic for the stealthiness of a cyberattack to augment traditional threat models. The methodology presented here will help in building the foundational principles of electric vehicle cybersecurity: a nascent but critical topic in the coming years.
\end{abstract}

\clearpage
Modern vehicles consist of a myriad of devices and systems ranging from safety-critical systems that control a vehicle's brakes to auxiliary components that adjust cabin temperature and wiper speed. While such components enhance the users' safety and comfort, they also render the vehicle's internal networks vulnerable to cyberattacks.  When these vulnerabilities are exploited, attackers can gain access to safety-critical systems like the brakes and transmission of the vehicle, as demonstrated by recent work.\cite{koscher2010autocyber, miller2015remote}

\noindent Alongside, another notable development in the automotive sector is the transition to electric vehicles (EVs) motivated by efforts to downscale tailpipe emissions.\cite{needell2016potential,kempton2016electric}  Widespread EV adoption is bottlenecked by limited driving range, battery pack cost, battery lifetime and safety issues associated with Li-ion batteries.\cite{yang2018predictive,needell2016potential,cano2018batteries,groger2015electromobility}  The battery pack also forms a significant fraction of the total cost of the electric vehicle ($\sim$20\% of the cost).\cite{safari2018battery,schmuch2018performance}  From the standpoint of automotive cybersecurity, while the primary focus is on immediate safety concerns, EV battery packs present several new vulnerabilities related to the bottlenecks mentioned previously. It is crucial to explore the cybersecurity aspects of battery packs to inform and improve future work in EV cybersecurity.\cite{pratt}


\begin{figure}
\centering
\includegraphics[width=0.9\linewidth]{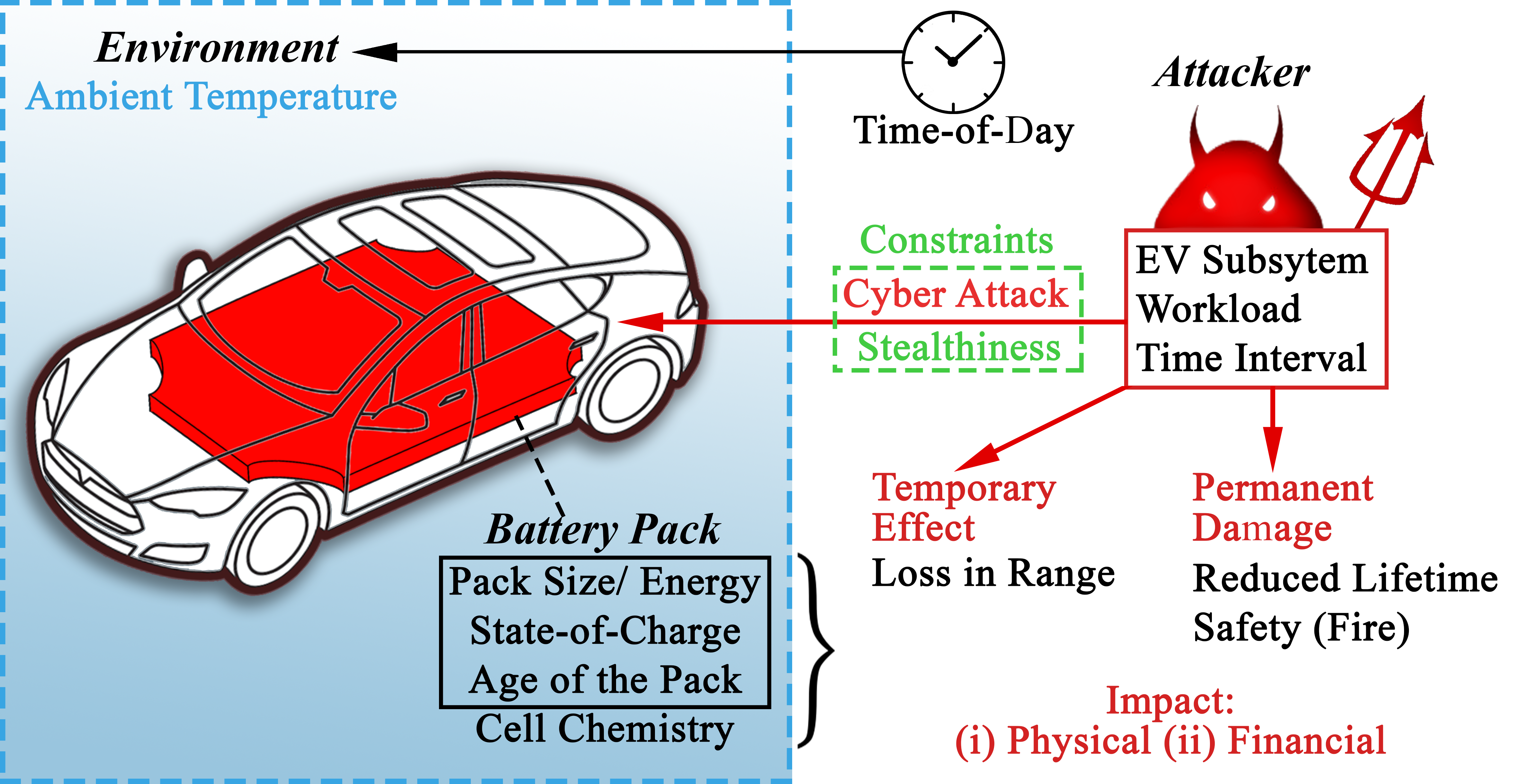}
\caption{A pictorial illustration of various potential attack scenarios. The illustration enumerates all the variables that need to be considered for analyzing the impact of cyberattacks, which could cause (i) Physical and/or (ii) Financial impact.  The attacker could utilize auxiliary components within the EV or EV charging systems using the attacker's control dimensions. Cyberattacks can cause either temporary effects or permanent damage.  The only environmental state variable of relevance in this scenario is the ambient temperature.  The different variables that define the state of the battery pack influence the magnitude of impact due to the cyberattack.
(The automobile outline illustration is published with permission from Chris Philpot.)}\label{fig-1}
\end{figure}

\noindent One of the challenges in assessing systems involving batteries is to accurately analyze the complex molecular-scale processes occurring inside a closed system.\cite{wang2002computational,ramadesigan2012modeling,dubarry2009} A practical battery system stores a fixed amount of energy via reversible electrochemical reactions. During normal operation, several unwanted side reactions also occur, which eventually degrade the battery's ability to store energy and thus, reduce the lifetime.\cite{christensen2,santhanagopalan2006review,yang2017modeling,safari2010life1,safari2010life2,ramadesigan2011parameter,peled2017sei} Further, from a safety standpoint, batteries have a specified set of conditions for safe operation, outside which the risk of thermal events increases. Cyberattacks may compromise driving range by draining energy via higher loads, reduce lifetime by enhancing side reactions, and compromise safety by pushing the operating conditions to unsafe limits, as illustrated in \ref{fig-1}.

\noindent \textit{Attack Scenarios:} The attack scenarios for an EV are centered around an attacker who aims to cause either physical or financial losses through cyberattacks. Modern vehicles, including EVs, contain several devices called Electronic Control Units (ECUs) that are responsible for a majority of vehicle's functions.  ECUs gather sensor inputs and actuate mechanical and electromechanincal components within the vehicle.\cite{koscher2010autocyber} Recent efforts that have demonstrated vulnerabilities in automotive networks, have primarily examined vehicles that employ the Controller Area Network (CAN) communication protocol.  CAN is the prevailing standard for intra-vehicle communication due to low cost and robustness; however, there are many CAN exploits.\cite{koscher2010autocyber} An attacker can gain access to the vehicle's CAN networks via direct physical access \cite{koscher2010autocyber, checkoway2011autocyber} or the remote exploitation of an ECU with existing direct access.\cite{miller2015remote} If an attacker aims to cause financial impact, one attack trajectory could be reducing the lifetime of the battery pack by enhancing the rate of degradation. In terms of physical damage, cyberattacks could increase the risk of thermal runaway where the attacker can overcharge or overdischarge the battery pack which are achieved through attacks on the battery management systems (BMS) when used in conjunction with other parasitic loads like auxiliary component loads. 

\noindent In this work, we develop a physics-driven approach which uses an experimentally validated battery model\cite{kalupson2013autolion} within a vehicle dynamics model to simulate the operation of an EV.\cite{sripad2017evaluation,sripad2017performance} We also explore new concepts like `stealthiness of attacks' and the trade-offs between stealthiness of attack and extent of damage from the attack. Using this framework, we quantify the impact of cyberattacks in different scenarios. We analyze either financial and physical losses incurred through either: (i) permanent damage,  defined as a change in the state of system that is irreversible, for example, irreversible capacity loss in a battery pack and (ii) temporary damage, defined as a change in the state of system that is (mostly) reversible, for e.g., reduction in state-of-charge which can be recovered by re-charging.



\textit{Permanent Damage:} As we stated previously, cyberattacks can accelerate cell degradation and shorten the lifetime of the battery pack.  Experimental demonstration of degradation is typically indirect as batteries are closed systems and measuring the internal states of the batteries is extremely difficult.\cite{fathi2014ultra,smith2011high} Thus, a validated physics-based model that can track the internal states of a battery packs provides a convincing means to demonstrate permanent damage due to cyberattacks.

\noindent Among the different mechanisms that cause cell degradation, two main processes of interest are: (i) growth of the solid-electrolyte interphase (SEI) layer at the graphite anode and (ii) lithium plating.\cite{christensen2,safari2010life1,safari2010life2,peled2017sei,safari2009multimodal,yang2017modeling} The SEI layer grows as a result of solvent reduction at the anode-electrolyte interface and consumes Li$^+$ ions, thereby causing a decrease in the amount of active Li$^+$ ions available and a reduction in capacity.  Plating of Lithium at the anode similarly leads to a loss in capacity along with an increase in the risk of internal shorts which could lead to catastrophic safety issues.\cite{abada2016safety,deng2018safety,deng18}

\noindent The permanent damage due to a cyberattack can be quantified using the rise in the internal resistance of the cell. The rise in the internal resistance is estimated using the increase in the thickness of the SEI layer.\cite{lawder,safari2009multimodal} The extent of Li-plating is controlled by the electrochemical potential for lithium deposition or the `Li-plating potential'.\cite{yang2017modeling}  The EV battery pack end-of-life is characterized by degradation in capacity to 80\% of the initial capacity.\cite{wood2011eveol,saxena2015quantifying}  We define the usable 20\% of the capacity as the `vital capacity' of the battery pack. A parametric study of the effect of different variables on degradation is compiled in the \textit{Supplementary Information}.

\begin{figure}[!ht]
\centering
\includegraphics[width=0.5\linewidth]{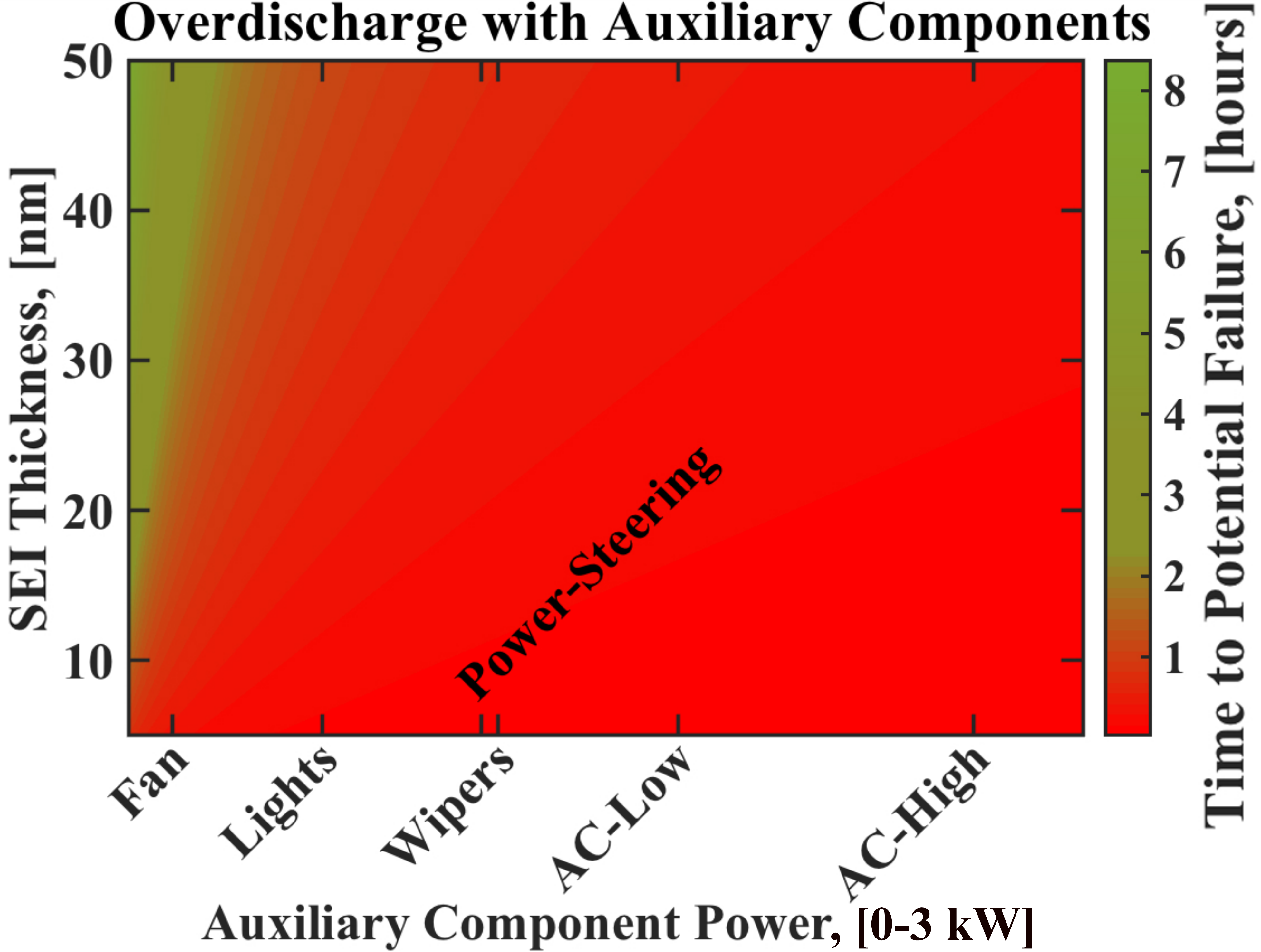}
\caption{A compromised battery management system, is vulnerable to attacks that override the lower voltage cutoff which can overdischarge the pack. During overdischarge, one of the initial steps is the decomposition of the Li-ion containing SEI layer which is followed by the dissolution of copper ions from the current collectors, with the possibility of internal shorts and other safety events. The estimated time to the onset of copper dissolution occurs during overdischarge is shown above for the cells based on NCA (Ni$\mathrm{_{0.8}}$Co$\mathrm{_{0.15}}$Al$\mathrm{_{0.05}}$O$\mathrm{_2}$) cathode and Graphite anode. For components with the power consumption equivalent to lights ($\sim$200 W), the time to onset of copper dissolution is under 2 hours while components with a high power consumption like air-conditioning have a timescale of less than an hour.}\label{fig-od}
\end{figure}

\textit{Compromised Battery Management Systems, Overdischarge:} When a BMS is compromised, an override of the lower cut-off voltage is possible.\cite{lee2005intelligent}  An attack on an EV with a depleted battery pack and compromised BMS can lead to overdischarge through energy-intensive auxiliary components. In terms of such cyberattacks occurring on an EV with the depleted battery pack, the idea of using wake-up functions as attacks has been demonstrated recently.\cite{cho2018killed} Such attacks could be followed by auxiliary component attacks discussed in this work, to overdischarge the cells.  During overdischarge, the initial stages involve the decomposition of the SEI layer which is composed of Lithium containing compounds and subsequently Copper dissolution from the current collector begins.\cite{guo2016mechanism} The dissolved Copper ions eventually lead to deposition of metallic Copper and potential internal shorts. The time for potential failure can be estimated using the time required for the decomposition of the SEI layer during the cyberattack, as shown in (Fig. \ref{fig-od}). The cells shown in (Fig. \ref{fig-od}) correspond to that of a 100 kWh battery pack based on NCA

(Ni$\mathrm{_{0.8}}$Co$\mathrm{_{0.15}}$Al$\mathrm{_{0.05}}$O$\mathrm{_2}$) cathode and Graphite anode. The thickness of the SEI layer is a function of the age of the battery pack where 50nm is assumed to be equivalent to a battery pack aged over 2 years, however, the thickness would change with the vehicle operating conditions.  We observe that attacks that involve components with an energy consumption rate of over 200W, the timescale for the complete decomposition of the SEI layer and potential failure is under 2 hours. While the consequences of overdischarge in Li-ion batteries depend on the kind of materials used in the cells, the impact could range from the loss of energy through the internal short to thermal and safety events as well.\cite{lee2005intelligent,guo2016mechanism}

\begin{figure}[!ht]
\centering
\includegraphics[width=0.95\linewidth]{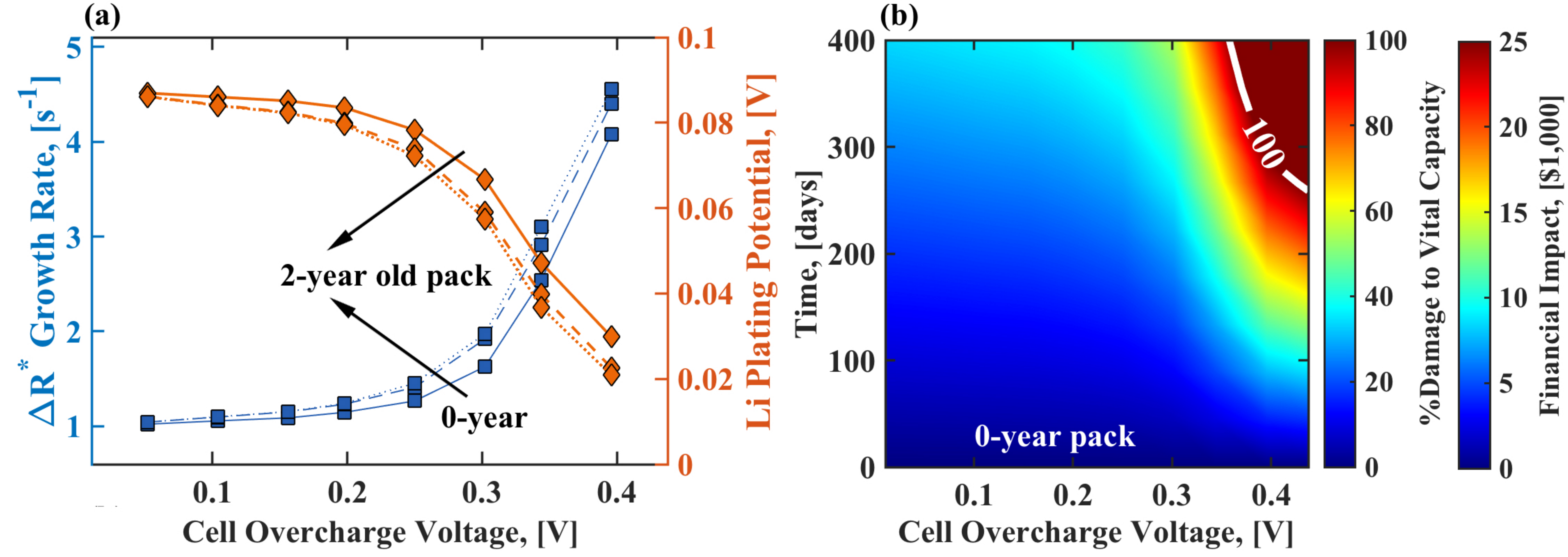}
\caption{The impact of cyberattacks on charging systems specifically aimed at overcharging the battery pack is summarized here.(a) The attacks studied here lasts for one hour after charging. We can study the increase in the SEI growth rate and $\mathrm{\Delta R^*}$.(b) This increase along with Li-plating translates to capacity fade and could shorten the lifetime to about 200 days at an overcharge voltage of about 0.4V. The reduction in the Li-plating potential due to overcharge in (Fig. \ref{fig-3}a), provides a metric to quantify the risk of developing internal shorts due to lithium plating.\cite{buggaplating}  When the same attacks are performed on older packs, we observe that the $\mathrm{\Delta R^*}$ growth rate increases while the Li-plating potential decreases, both of which are detrimental to the state-of-health of the battery pack.}\label{fig-3}
\end{figure}

\textit{Compromised Battery Management Systems, Overcharge:}
A compromised BMS can modify the upper cut-off voltage.\cite{lelie2018battery}  The pack can then be charged at a voltage higher than the normal charging voltage (manufacturer specific upper voltage cutoff) leading to overcharging.  Within a constant current-constant voltage protocol,\cite{zhang2006effect} an increase in the charging current would lead to an increased rate of degradation which is an extension of the previously mentioned parametric analysis on the discharge rate of the battery pack.  However, overcharging the battery pack leads to various other issues shown in (Fig. \ref{fig-3}).

\noindent In (Fig. \ref{fig-3}a), we observe a super-linear rise in the growth rate of $\mathrm{\Delta R^{*}}$ as the overvoltage per cell increases in a fresh cell.  The charging system cyberattack simulated spans a duration of one hour after charging similar to the auxiliary component cyberattacks. However, the consequent damage caused to the battery pack in terms of capacity fade, as shown in (Fig. \ref{fig-3}b), is enormous. At a cell overvoltage of 0.4V, we observe that the pack reaches its end-of-life or 100\% damage to vital capacity in about 200 days. This could result in significant financial impact as shown in (Fig. \ref{fig-3}b) where we estimate the monetary value of the loss of capacity for a 100kWh battery pack assuming the cost of battery packs of about \$200/kWh.\cite{kittner2017energy} (Fig. \ref{fig-3}a) also shows the decrease in the Li-plating potential which implies that lithium would plate more readily at higher overvoltage. Over time, such attacks could lead to an increased amount of Li-plating which could have safety implications resulting in physical impact including thermal events and fire.\cite{abada2016safety,wang2012safety}

\textit{Compromised Auxiliary Components:} Compromised auxiliary components effectively act as parasitic loads. Quantifying the impact of such attacks requires a close examination of different operating and environmental variables.  The variations in each of the state variables like temperature, state-of-charge (SOC), pack size, age of the pack, etc. and the set of variables that defines a given auxiliary component attack workload\cite{load1,load2,load3} affects the degradation in vital capacity in a different manner. A parametric analysis of all the variables, exploring the effect of each variable, similar to other studies\cite{safari2010life1,safari2010life2} reveals that the damage to vital capacity increases with the temperature by following the Arrehenius relationship which implies that cyberattacks conducted at higher ambient temperature would cause greater impact. Damage to vital capacity also increases with the State-of-Charge of the battery pack which suggests that attacks on fully charged battery packs would cause more damage. As the age of the pack increases, the damage caused by a fixed load in the same conditions decreases. The damage to vital capacity is seen to be a sub-linear function of the total time of attack, characteristic of a diffusion-limited process. Further, damage to vital capacity increases linearly with an increase in the cumulative energy consumption of the load, a phenomenon which has been covered previous studies on capacity fade.\cite{safari2010life1,safari2010life2}

\begin{figure}
\centering
\includegraphics[width=0.5\linewidth]{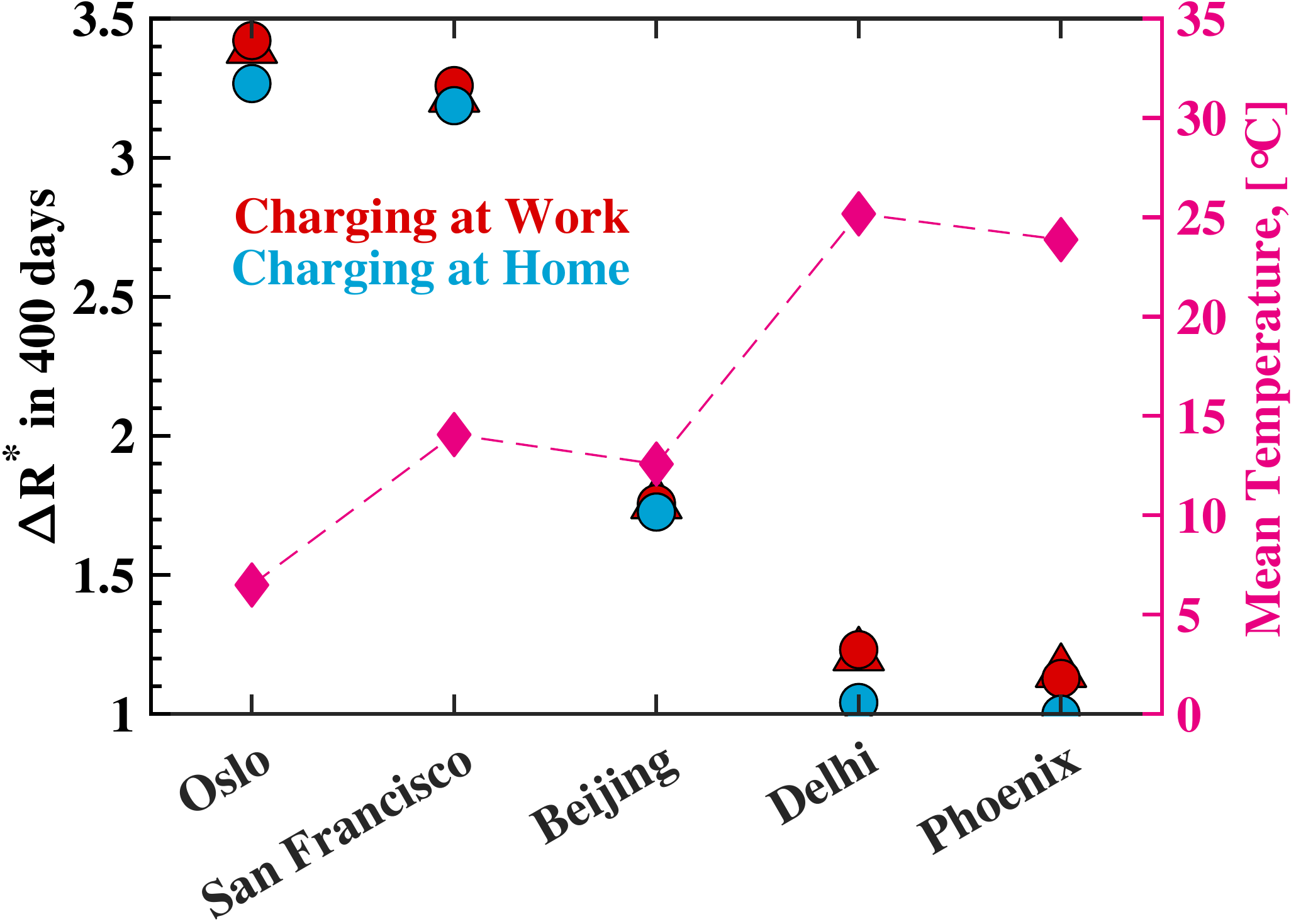}
\caption{We can study the impact of auxiliary component cyberattacks here, based on the results for simulations equivalent to $\mathrm{\sim}$400 days. $\mathrm{\Delta R^{*}}$ represents the increase in the resistance of the cell due to the cyberattack when compared to the baseline scenario. The triangular markers indicate situations where the vehicle is attacked twice in a day. In such cases, the average $\mathrm{\Delta R^{*}}$ of the two attacks is shown, while the circular markers represent the cases with one attack over the day.}\label{fig-2}
\end{figure}

\noindent Following the insights from the parametric analysis, we infer that attacks which comprise of energy intensive auxiliary components when engaged after a new battery pack is fully charged cause the most damage. We design the attack scenarios accordingly.  We consider two types of EV users based on charging behavior, either charging at `Home' or charging at `Home' and at `Work'.  The sample attack workload spans a duration of one hour and is based on the combination of A/C at high power along with Lights, Power-Steering and Wipers.  We analyze the cases where these users are located in Oslo, San Francisco, Beijing, Delhi and Phoenix which serve as proxies for the environment state variable of temperature and are chosen to represent a wide range of temperature conditions. In order to analyze the impact of auxiliary component cyberattacks, we use $\mathrm{\Delta R}$, a quantity which represents the increase in internal resistance of the cell compared to a cell which has not been subjected to the attack workloads. $\mathrm{\Delta R}$ essentially provides information on the effectiveness of the cyberattack.  We calculate $\mathrm{\Delta R}$ after 400 days for each case using the following relationship,
\begin{equation}
\centering
\mathrm{\Delta R = \dfrac{R_{SEI}^{A} - R_{SEI}^{B}}{R_{SEI}^{B}}},
\label{eq-deltar}
\end{equation}
\noindent where $\mathrm{R_{SEI}}$ is the resistance due to the SEI layer, and `A' and `B'  represent the attack and baseline scenario. For the quantities reported in (Fig. \ref{fig-2}), $\mathrm{\Delta R^{*}}$ values are obtained by normalizing all the $\mathrm{\Delta R}$ values with the minimum value in a given set which facilitates the comparison of values within the set.



\noindent In (Fig. \ref{fig-2}), the rise in $\mathrm{\Delta R^{*}}$ is the most for Oslo, which has the lowest average ambient temperature.  While an increase in ambient temperature causes an increase in the thickness of the SEI layer, the resistance due to the formation of SEI layer impedes further growth.\cite{christensen}  This phenomenon leads to the fact that places like Phoenix, where the ambient temperature is high, already feature a substantial SEI layer thickness, thereby minimizing any additional damage to vital capacity due to the attack workload.  However, it is worth highlighting that the average resistance due to the SEI film formed is higher in warmer regions compared to colder regions.  The two cases of charging, at Home and at Work, do not show any substantial difference, although, if the EV is charged in both locations, then we have two separate time windows for attack.

\textit{Temporary Damage:} With compromised auxiliary components, attack workloads can cause a depletion of energy contained in the battery pack, thereby a reduction in available driving range. This damage can be reversed by charging the battery pack.  However, such attacks can play into the well-known issue of `range anxiety'.\cite{needell2016potential}  For some vehicles, with battery packs <40kWh battery packs, up to 20\% of the available range could be depleted in under one hour with energy intensive attack workloads which include combinations of auxiliary components as discussed previously. Such attacks which engage several components at the same time will be more energy intensive compared to single components, however, such attacks might be easier to detect for the user which is discussed in the subsequent sections. 

\textit{Stealthiness of Attack:} An important constraint on a cyberattack is the likelihood of it getting detected.  In the case of auxiliary components, the detection is by the user and hence it is difficult to develop a quantitative metric for the same.  However, in order to provide a basic overview of the issue, we develop a qualitative understanding using three scenarios, namely, `parked', `stationary' (at rest within driving operation) and `driving'.  A summary of the stealth of an attack involving a given auxiliary component is given in (Tab. \ref{table}).  Such a metric is heuristic but it provides a calibration for the components that are more likely to be targeted based on the attacker's perspective.  An auxiliary component that involves a high stealthiness of attack and is also energy intensive would naturally be targeted often.
\begin{table}
\centering
\scriptsize
\setlength{\tabcolsep}{1.5mm}
\caption{Stealthiness of attack, a qualitative metric used by attackers to reduce the chance of detection.}
\begin{tabular}{l|p{35pt}|p{35pt}|p{35pt}}
\multirow{2}{*}{Auxiliary Component} & \multicolumn{3}{c}{Stealthiness of Attack} \\ \cline{2-4}
    & Parked & Stationary & Driving\\
\hline
A/C-High & High & Low & Medium \\
A/C-Low & High & High & High \\
Power Steering & N/A & High & High \\
Lights & High & Low & Medium \\
Fan & High & Low & Medium \\
Wipers & Medium & Very Low & Very Low \\
\textit{Combinations} & High & Low & Low\\
\hline
\end{tabular}
\label{table}
\end{table}

\textit{Rowhammer Attack:} Rowhammer style attacks\cite{kim2014flipping} have been demonstrated previously where targeted workloads on memory systems were generated to cause corruptions which can be used to launch further attacks. We observe an analogous case here with battery systems since battery pack is made up of several cells arranged in a matrix involving a series-parallel configuration.  This architecture is vulnerable to `rowhammer' attacks since individual strings or cells within battery packs could be targeted through a compromised battery management system and the damage to individual strings or cells is magnified.  Each of the cyberattack scenarios we have considered, like attacks on auxiliary components, overcharge, and overdischarge could be orchestrated as rowhammer attacks.  We previously discussed the various factors due to which the damage to the battery pack increases with a reduction in pack size for the same workload which is especially relevant to rowhammer attacks.  Such attacks could not only shorten the lifetime of the targeted subset of the battery pack but could also lead to issues related to instabilities due to the isolation of strings within the battery pack.


\noindent We have discussed the potential physical and financial impact due to cyberattacks on EVs and EV subsystems.  We identify simple but effective cyberattacks on auxiliary components that can temporarily drain the battery pack up to 20\% per hour.  Furthermore, we analyze attacks could lead to a deterioration in the power capability due to an increase in the cell resistance.  We use a metric which is equivalent to the `normalized resistance increase', which can be used to quantify the extent of performance reduction. We find that normalized resistance increase is generally higher for colder regions.  We find that cyberattacks on auxiliary components launched after the pack is completely charged (i.e. high state-of-charge) leads to more damage.  The cell resistance increase, largely due to the formation of a solid-electrolyte-interphase (SEI) layer, follows a sublinear relationship with time.  This results in a new pack being more vulnerable than an aged pack to cyberattacks on auxiliary components.  Compromised battery management systems expose the pack to two kinds of attacks, (i) Overdischarge and (ii) Overcharge.  Overdischarge attacks which override the lower cutoff voltage of the pack could lead to the complete decomposition of the SEI layer in under two hour thorough auxiliary components with a power rating of over 200W. The decomposition of the SEI is followed by the dissolution of Copper ions which could eventually lead to internal shorts and potential safety events.  Cyberattacks launched during charging through the compromise of the voltage regulator could lead to an overcharge of the cells, which in some cases could even lead to physical safety issues (e.g. fire).  Further, this could lead to a new pack being depleted to 80\% of its initial capacity (end-of-life for an EV battery) in less than a year.  Finally, a compromise of the battery management system could lead to novel ``rowhammer"-style attacks (attacking a string of cells), which could damage a subset of cells in a short time span.   We believe that the results presented here will inform the development of robust detection and prevention systems and provide a much-needed rational design approach for electric vehicle automotive security.\cite{khalid}


\begin{addendum}
 \item S. S., S. K., V.S. and V. V. gratefully acknowledge support from Technologies for Safe and Efficient Transportation University Transportation Center. V.S. and V. V. gratefully acknowledges support from the Pennsylvania Infrastructure Technology Alliance, a partnership of Carnegie Mellon, Lehigh University and the Commonwealth of Pennsylvania's Department of Community and Economic Development (DCED).
 \item[Competing Interests] The authors declare that they have no
competing financial interests.
 \item[Correspondence] Correspondence and requests for materials
should be addressed to V.V.\\~(email: venkvis@cmu.edu).
\end{addendum}

\begin{methods}
\subsection{Battery Pack Simulations.} The description of the system of equations for the multiphysics battery model\cite{kalupson2013autolion, yang2017modeling} can be found elsewhere. Cells constructed based on this modelling framework are assembled into a battery pack model. The baseline load profile for the vehicle is based on Urban Dynamometer Driving Schedule for 50 miles per day, along with a constant current-constant voltage (CC-CV) charging protocol with a peak power of a level-1 charger.  The attack workloads are implemented within this daily load profile.  The daily load profiles are repeated to simulate the operation over a long period.  The battery model is a thermally coupled model,\cite{kalupson2013autolion} and the ambient temperature conditions are implemented within the same simulation framework, described in detail in the Supporting Information.

\end{methods}

\noindent \textbf{Supporting Information (SI):}
Supporting Information contains details of the parametric study for battery degradation and other information on the battery modeling undertaken for the study.


\bibliographystyle{naturemag}
\bibliography{pnas-bib}



\section*{Supplementary material}

# Supplementary Information: Vulnerabilities of Electric Vehicle Battery Packs to Cyberattacks

Shashank Sripad,[a] Sekar Kulandaivel,[b] Vikram Pande,[a] Vyas Sekar[b] and Venkatasubramanian Viswanathan[a*]

[a]Department of Mechanical Engineering, Carnegie Mellon University, Pittsburgh, Pennsylvania, 15213, USA.

[b]Department of Electrical and Computer Engineering, Carnegie Mellon University, Pittsburgh, Pennsylvania, 15213, USA.

*Corresponding author, Email: venkvis@cmu.edu

## Battery Modeling

These equations used for the battery modeling are summarized by Fang et al.,[1] and Kalupson et al.,[2] using Eqns. (1-5) which describe the 1-D transport model for the species and the charge coupled with a lumped thermal model. The solid phase charge conservation is given by

$$\nabla . (\sigma^{\text{eff}} \nabla \Phi_s) = j^{\text{Li}}, \tag{1}$$

where $\sigma^{\text{eff}}$ is the effective electronic conductivity and $\Phi_s$ is the potential of the solid phase. The electrolyte phase charge conservation is given by

$$\nabla . (k^{\text{eff}} \nabla \Phi_e) + \nabla . (k_D^{\text{eff}} \nabla \log(c_e)) = -j^{\text{Li}}, \tag{2}$$

where $k^{\text{eff}}$ is the effective ionic conductivity, $\Phi_e$ is the potential of the electrolyte phase, $k_D^{\text{eff}}$ is the conductivity and $c_e$ is volume averaged Li concentration in the electrolyte phase. The conservation of species in the electrolyte phase is given by

$$\frac{\partial(\varepsilon_e c_e)}{\partial t} = \nabla . (D_e^{\text{eff}} \nabla c_e) + \frac{(1 - t_+^0)}{F} j^{\text{Li}}, \tag{3}$$

where $\varepsilon_e$ is the volume fraction, $D^{\text{eff}}$ is the effective diffusion coefficient, $t_+^0$ is the transference number, and F is the Faraday constant. The species conservation in solid phase is given by

$$\frac{\partial c_s}{\partial t} = \frac{D_s}{r^2} \frac{\partial}{\partial r}\left(r^2 \frac{\partial c_s}{\partial r}\right), \tag{4}$$

where $c_s$ is the concentration of Li in the solid phase, $D_s$ is the diffusion coefficient in the solid phase, and 'r' represents the radius of the particles of active material. The energy

balance is represented as a lumped thermal model given by

$$\frac{\partial(\rho C_p T)}{\partial t} = (q_r + q_j + q_c + q_e)A_{cell} + h_{conv}A_s(k\nabla T), \tag{5}$$

which accounts for $q_r$ reaction heat, $q_j$ the joule heating, $q_c$ the heating due to contact resistance between the current collector and electrode materials, and $q_e$ the entropic heating. The last term represents the heat dissipation, where $h_{conv}$ is the coefficient of heat dissipation and $A_s$ is the cell external surface area.

## SEI layer and Degradation Processes

The parasitic reactions within the cell based on models that can be found elsewhere[3,4] are:

$$j_{SEI} = -k_{SEI} \cdot c_{sol}^s \cdot \exp\left[-\frac{\alpha_{c,SEI}.F}{RT}\cdot(\phi_s - \phi_e - I.R_{SEI} - U_{SEI})\right], \tag{6}$$

$$j_{PL} = -i_{o,PL} \cdot \exp\left[-\frac{\alpha_{c,PL}.F}{RT}\cdot(\phi_s - \phi_e - I.R_{SEI})\right], \tag{7}$$

where the side currents for each of the degradation processes for Solid-Electrolyte Interphase (SEI), ($j_{SEI}$), for the Lithium plating, ($j_{PL}$), and the last rate equation captures the Active Material Isolation along with the total current, (I). The other constants from the degradation sub-model are the rate constants ($k_{o,SEI} = 1 \times 10^{-12}$ m/s)[4], ($k_{AMI} = 2 \times 10^{-14}$ m/s) and the exchange current density, ($i_{o,PL} = 0.001$ A/m$^2$)[4]. The ($\alpha$'s) are the cathodic transfer coefficients. ($c_{sol}^s$), is the concentration of the solvent. The ($\phi$'s) are the potentials of the electrode and liquid phases. ($R_{film}$) is the resistance of the SEI layer.

The increase in the thickness of the SEI layer increases the internal resistance or impedance of the cell, thereby leading to a loss in the power capabilities.

$$\frac{d\delta_{SEI}}{dt} = -\frac{j_{SEI}}{2F}\frac{M_{SEI}}{\rho_{SEI}}, \tag{8}$$

where $M_{SEI}$ and $\rho_{SEI}$ are the molecular weight and the density of the SEI. The resistance due to the SEI is calculated using the effective conductivity of the electrolyte (solvent) through the SEI using:

$$R_{SEI} = \frac{\delta_{SEI}}{\kappa_{SEI}^{eff}}, \tag{9}$$

## Parametric Analysis

Throughout this study, we examine capacity fade considering a quantity defined as 'vital capacity' which is equivalent to 20% of the initial capacity of the battery pack. For EVs a degradation of 20% of the initial capacity or 100% of the vital capacity marks the end-of-life of the battery pack[5]. In order to examine the effect of each variable, we select a sample cyber attack defined by a specific workload on the pack conducted for a fixed duration of time. In

other words, we fix all the attacker's control dimensions and constraints. We also fix all other battery pack and environment state variables apart from the variable in consideration, and thereby examine permanent damage as a function of the given variable alone. For examining permanent damage, the first variable considered for analysis is the ambient temperature. The damage to vital capacity is observed to be an exponentially increasing function of ambient temperature and follows the Arrhenius relationship. The parametric analysis summarized in (Fig. S1-S4) are the results of battery pack simulations for attack workloads of one hour comprising of A/C, Lights, Power-Steering and Wipers, for 400 attack-charge cycles. The Constant Current Constant Voltage (CC-CV) charging protocol is followed in each of the simulations.

The effect of ambient temperature on the damage to vital capacity is illustrated in (Fig. S1). The damage to vital capacity follows the Arrhenius equation and within the same analysis, on comparing different workload, we can observe the damage to vital capacity increases linearly as a function of the average power of the workload.

The SOC along with temperature are studied to see their influence on the capacity fade for battery packs of different age, namely 0, 1, and 2-year-old packs as visualized in (Fig. S2). We observe that the damage to vital capacity increases with increase in SOC, in other words, if a cyber attack is conducted on a fully-charged battery pack or a battery pack which is placed in a higher ambient temperature, then the resultant impact would be much greater.

On comparing the SOC-Temperature results for packs of different age, we observe an interesting trend that older packs show a much lower damage to vital capacity with the same specified cyber attack. This can be explained from the fact that the SEI layer modeled shows a growth rate which is a sub-linear function of time, and hence the capacity fade or damage to vital capacity decreases for a battery which has already aged for a certain duration.

The effect of total time-of-attack on damage to vital capacity can be seen in (Fig. S3), where we see that the damage is a sub-linear relationship of time. The damage to vital capacity also increases linearly with a reduction in the battery pack size. This could be explained as capacity fade being a linear function of the depth-of-discharge, as seen in (Fig. S4) since a smaller battery pack would have a larger depth of discharge for the same auxiliary component or workload considered, and this corroborates the findings of other studies[6].

## Disabling Battery Cooling Fan

To demonstrate the feasibility of causing a thermal runaway event, we have discovered a particular CAN message that disables the battery cooling fan for a 2009 Toyota Prius. To discover this message, we systematically injected a CAN message with each possible ID value ranging from 0x0 to 0x6FF, or the safe range for message injection[7]. For each injection, we flood the network with a particular ID and randomly modify the data values until we notice a physical response in the vehicle. We do acknowledge that the Prius is a Hybrid vehicle and that it is an older model car; however, modern EVs also have their Battery ECU connected to the CAN bus and a newer model car is much more likely to have minor systems, such as the cooling fan, controlled by CAN messages. If we assume that the attacker is a strong attacker (as discussed in our threat model) who can re-flash the Battery ECU, then the attacker can simply write code that ignores the battery temperature. With the combination

of disabling the battery fan and forcing the ECU to ignore battery temperature, it is clear how an attacker can cause a safety event.

## Fraction of Range Reduction and the Effect of Temperature

The energy consumption per unit distance for each EV is collected from the data of the U.S. Environmental Protection Agency. The ambient temperature has a significant impact on the rated range and the use of auxiliary components like heaters and air-conditioners at low or high ambient temperatures causes an additional loss in range. The SOC is could have an effect on the discharge efficiency of the battery pack and hence would affect the energy consumption. An aged battery pack would have a lower pack energy, and hence a lower overall rated range. For more details on the impact of each of these variables on short-term impact. The loss in range is determined by the energy consumed by the auxiliary components and the energy consumption of the vehicle at the time instant of the attack. In effect, the fraction of range lost or the ratio of range lost due to a cyberattack and the rated range of the vehicle would be independent of the temperature.

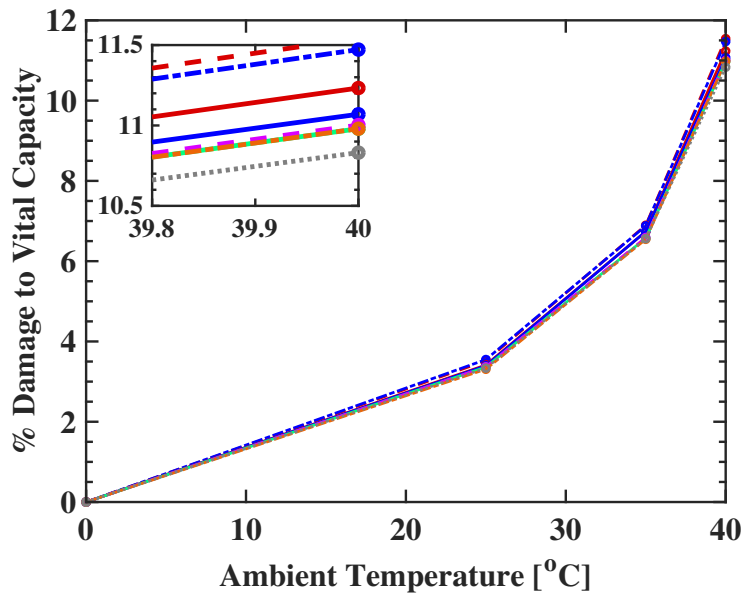

**Figure S1:** Damage to Vital Capacity as a function of ambient temperature for various auxiliary components, where we observe the Arrhenius type relationship. Each attack workload is run for a time duration of one hour and then subjected to charging with the pack SOC of about 0.7.

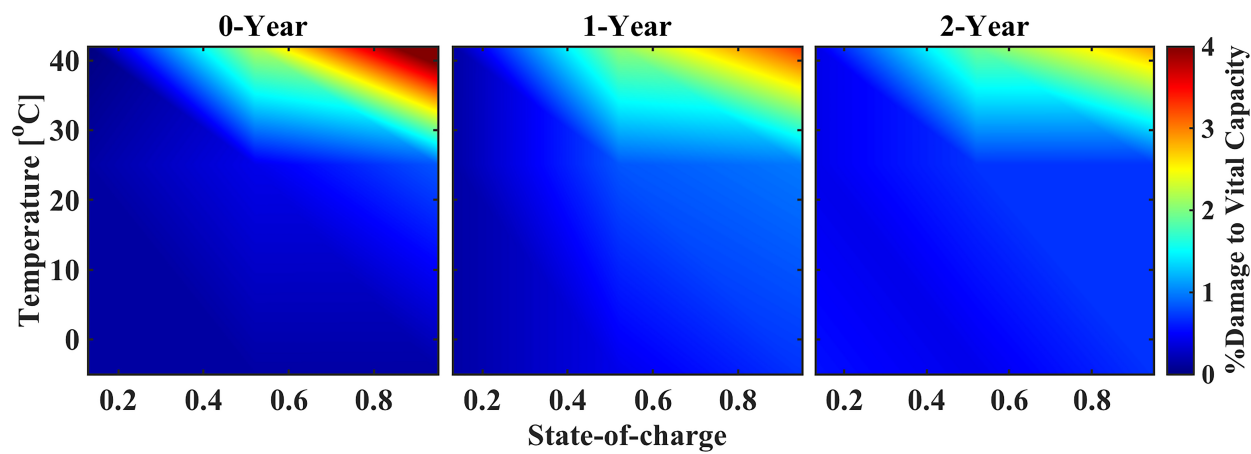

**Figure S2:** Bi-variate analysis of the Damage as a function of the SOC and ambient temperature conducted for battery packs of different age (0,1,2 years). The attack workloads are fixed for a time duration of one hour and then subjected to charging.

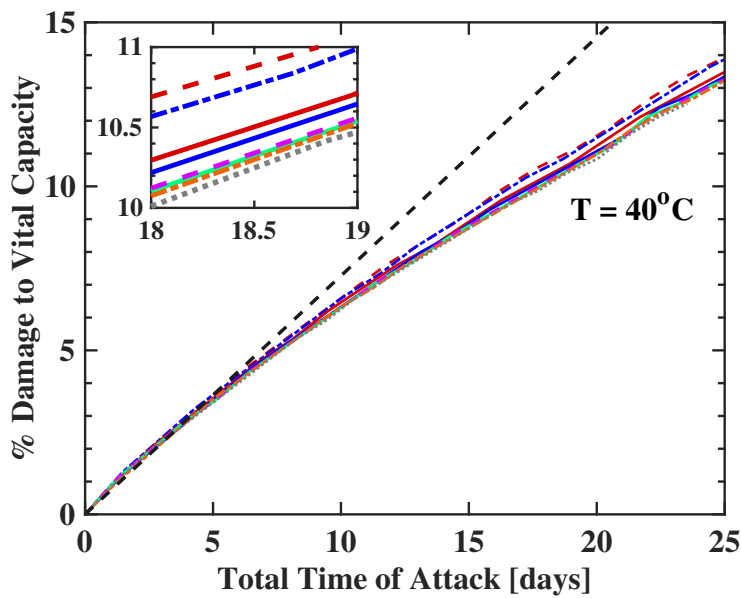

**Figure S3:** Damage to vital capacity studied as a function of the total time of attack for various auxiliary components which shows the sub-linear relationship. Each attack workload has a duration of one hour, and the pack is then subject to charging. The initial SOC of the pack is maintained at 0.7. The inset shows the different workloads that have been simulated.

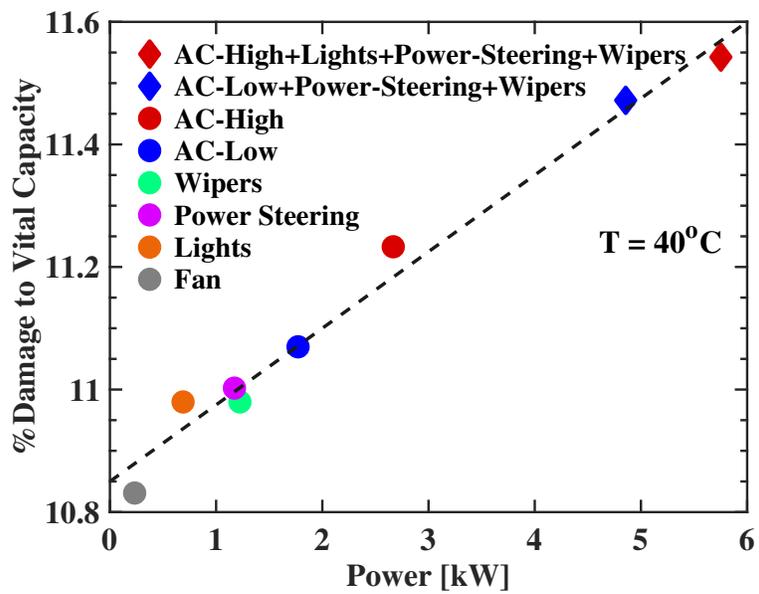

**Figure S4:** Damage to vital capacity studied as a function of the energy consumption of the auxiliary components at an ambient temperature of 40°C. The damage is calculated after 20 days of attack where the battery pack is subjected to the attack workloads of one over and charged after. The initial SOC of the pack is maintained at 0.7.

6

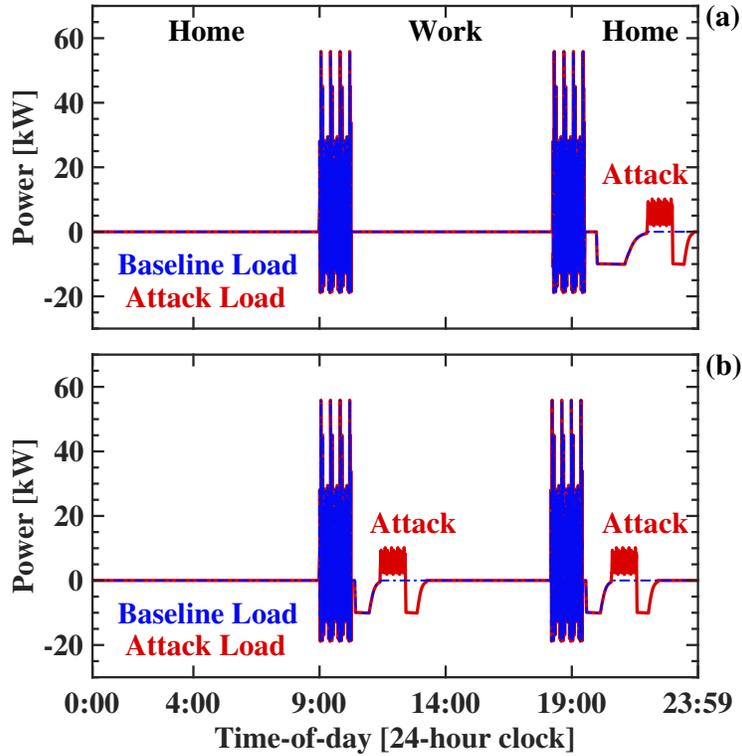

**Figure S5:** The daily power load on the Model S P100D which shows multiple scenarios. The power load shown in 'blue' represents the baseline scenario where we have no cyberattacks, and the 'red' represents the attack scenario. The attack is assumed to take place only during the time-window after charging based on the conclusions drawn from the our analyses above. The case where the user charges only at home is illustrated in (a) while (b) is a case where the user charges both at home and at work. The attack scenario in the work-charging case could have two distinct sub-cases where the attack is conducted either at one or both the time-windows after charging.

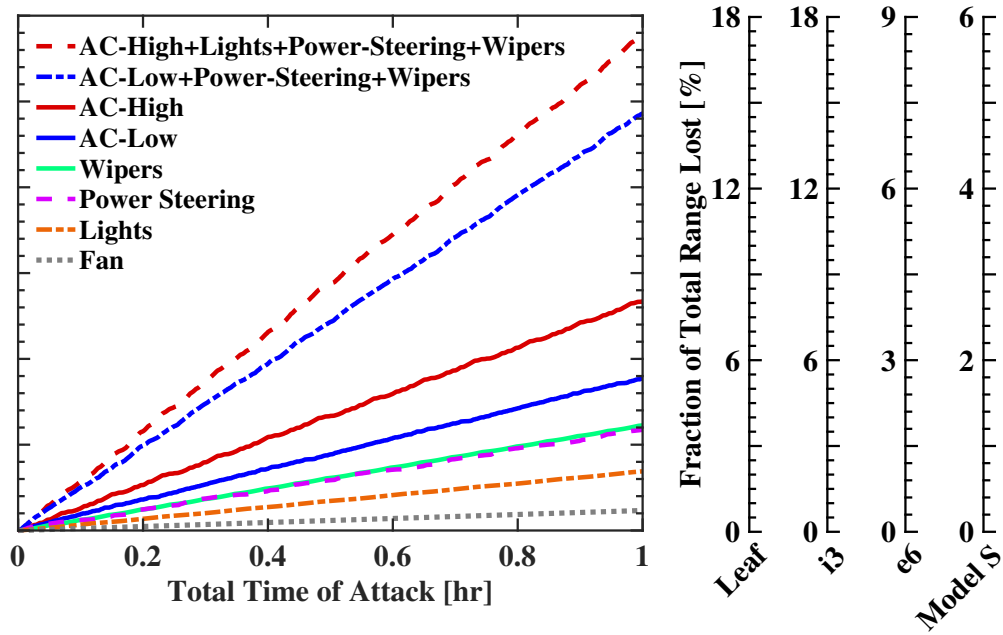

**Figure S6:** The fraction of rated range lost due to the cyberattack is determined using the energy consumption per unit distance of the electric vehicle. The energy consumption changes with temperature, where extreme (low of high) temperatures lead to an increased energy consumption. The quantity of range reduced is calculated using the same energy consumption as the one used to calculate the rated range, and hence the fraction of rated range lost is independent of the ambient temperature.

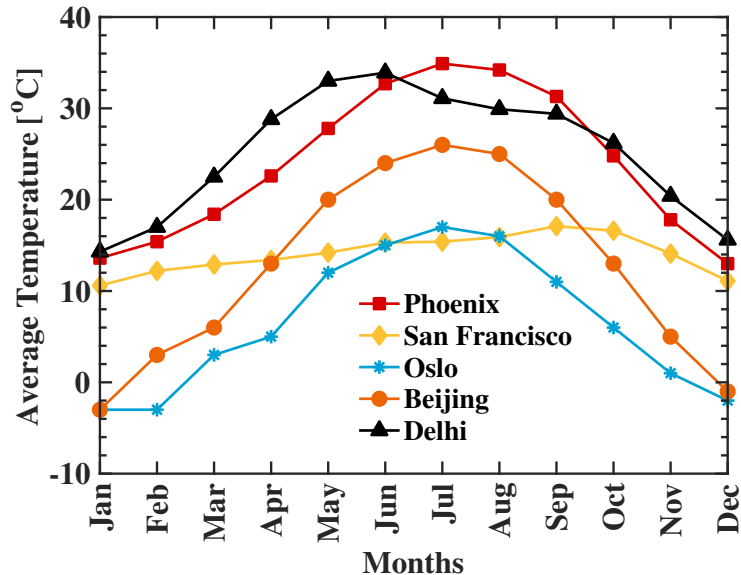

**Figure S7:** The temperature variations in different locations over the year used for the long-term damage studies. The ambient temperature data for each location is incorporated as an average over every week, and the effect of these changes in temperature can be seen in the SEI growth which is quantified in the model.

8



\end{document}